\documentclass[aps,preprint,groupedaddress]{revtex4}

\usepackage{graphics,epsfig}

\begin{document}

\title{Constraints on unparticle physics from the $gt\bar t$ anomalous coupling}
%\author{R. Martinez\\
%Departamento de F\'{\i}sica, Universidad Nacional, \\
%Bogot\'a, Colombia\\
%M.A. P\'erez \\
%Departamento de F\'{\i}sica, Cinvestav-IPN, Av. IPN\\
%2508, 07000 M\'exico, D.F., M\'exico\\
%O.A. Sampayo\\
%Departamento de F\'{\i}sica, Universidad Nacional del Mar de
%Plata\\
%FUNES 3350, 76000 Mar del Plata, Argentina}

\author{R. Mart\'{\i}nez}
\email{remartinez@unal.edu.co}
 \affiliation{Departamento de
F\'{\i}sica, Universidad Nacional, Bogot\'a, Colombia}
\author{M.A. P\'erez}
\email{mperez@fis.cinvestav.mx}
\affiliation{Departamento de
F\'{\i}sica, Cinvestav-IPN, Av. IPN 2508, 07000 M\'exico, D.F.,
M\'exico}
\author{O.A. Sampayo}
\email{sampayo@mdp.edu.ar}
\affiliation{Departamento de F\'{\i}sica,
Universidad Nacional de Mar del Plata FUNES 3350, 76000 Mar del
Plata, Argentina}

\begin{abstract}
We study the impact of unparticle physics to the chromomagnetic
dipole moment (CMDM) of the top quark. We compute the effect induced
by unparticle operators of scalar and vector nature coupled to
fermions on the CMDM. We find that this dipole moment is sensitive
to the scale dimensions $d_u$ of the unparticle and the new
couplings of the respective effective operators. Using the bounds
imposed on the CMDM by low-energy precision and Tevatron
measurements we derive indirect limits on the unparticle parameter
space.  In particular, we find that the scalar-unparticle operator
contribution fulfills both constraints for most of the unparticle
parameter space, while the low-energy precision bound on the CMDM
excludes a vector-unparticle contribution for low values of
respective scale dimension $d_u$.
\end{abstract}

\maketitle

One of the main goals of the CERN Large Hadron Collider (LHC) is to
study the properties of the top quark with significant precision
since more than 80 millions of top-quark pairs will be produced with
an integrated luminosity of $100$ fb$^{-1}$ \cite{uno}. In
particular, the interest in the study of its chromomagnetic dipole
moment (CMDM) $\Delta\kappa$ has increased recently since it was
realized that the presence of no-standard model couplings could lead
to modifications in the total and differential cross sections of
top-quark pairs  at hadron colliders \cite{dos}-\cite{ocho}. The
effect of $\Delta\kappa \neq 0$ has been examined in the Standard
Model, two Higgs doublets models (THDM), top-color assisted
Technicolor (TC2) models, 331 models, extended models with a single
extra dimension \cite{nueve}, and in the littlest Higgs model with
T-parity (LHT) \cite {diez}. In all these cases, it has been found
that their predictions for $\Delta\kappa$ are consistent with the
constraints imposed on $\Delta\kappa$ by low-energy precision
measurements obtained from the $b\to s\gamma$ process
$\Delta\kappa=-0.01\pm 0.048$ at $95\%$ C.L.\cite{nueve2}, and by
the cross section measurements of the Tevatron for $t\bar t$ pairs,
$|\Delta\kappa|\leq 0.20$ \cite{once}. In the present letter, we
compute the contribution induced on $\Delta\kappa$ by scalar and
vector unparticle operators \cite{doce}. Assuming the known bounds
on $\Delta\kappa$ obtained from the $b\to s\gamma$
process\cite{nueve2, once}, we derive limits on the scalar- and
vector-like unparticle dimensions $(d_u)$ and the new scale
$\Lambda$ that characterizes these interactions. We find that these
limits are similar to those obtained from cosmological and
astrophysical processes \cite{trece}, CP-violating effects in B
decays \cite{trece2}, Z boson decays \cite{quince} and the Tevatron
measurements for $t\bar t$ production \cite{diecisiete}.

Scale invariance has been a powerful tool in several branches of
physics. Scale invariant field theories have been investigated also
extensively. In particular, Georgi \cite{doce} has proposed that a
scale invariant sector, with no trivial $IR$ fixed point and which
couple to the $SM$ fields, may appear much above the TeV energy
scale. Below this energy scale, this sector induces unparticle
operators ${\cal O_U}$ with non-integral scale dimensions $d_u$ that
in turn have a mass spectrum which looks like a $d_u$ number of
massless particles. The couplings of these unparticles to the SM
fields are described by the effective Lagrangian \cite {catorce,
dieciseis}
\begin{eqnarray}
{\cal L}_{eff}\!\!\! &\sim&\!\!\! \frac{\zeta_S}{\Lambda^{d_u-1}}
\bar{t} t {\cal O}_U+ \frac{i\zeta_A}{\Lambda^{d_u-1}} \bar{t}
\gamma_5 t {\cal O}_U
\nonumber \\
&+& \frac{C_V}{\Lambda^{d_u-1}} \bar{t} \gamma_\alpha t {\cal O}^\alpha_U+ \frac{C_A}{\Lambda^{d_u-1}}
\bar{t}\gamma_\alpha\gamma_5 t {\cal O}^\alpha_U ,
\end{eqnarray}
where $\Lambda$ is the energy scale at which scale invariance
emerges, the dimensionless coefficients $\zeta_{S,A}$ and
$C_{V,A}$ are of order $1$ and $t$ is the top-quark spinor. Since
the operators with lowest possible dimension have the most
powerful effect in the low energy effective theory, in Eq. (1) we
have included only the scalar and vector operators ${\cal O}_U$
and ${\cal O}^\alpha_U$, respectively, of the unparticles that
couple to the quark-top.

If the source of this new physics is at the TeV scale, it has
been pointed out \cite{tres} that the leading effect on the top
quark-gluon interaction may be parametrized by the chromomagnetic
dipole moment (CMDM) $\Delta\kappa$ of the top quark since this
is the lowest dimension CP-conserving operator arising from an
effective Lagrangian contributing to the gluon-top-quark coupling,

\begin{equation}
{\cal L}_5 = i(\Delta\kappa/2)(g_S/2m_t)\bar{t} \sigma_{\mu\nu}q^\nu
T^\alpha t G^{\mu,\alpha}
\end{equation}
where $g_S$ and $T^\alpha$ are the $SU(3)_c$
coupling and generators, respectively, and the gluon is on-shell.

We consider that the scalar and vector unparticle mediation is
responsible for the CMDM and we predict the appropriate range for
the free parameters appearing in the effective Lagrangian which
drive the unparticle-SM quark interactions. We will obtain that
$\Delta\kappa$ is strongly sensitive to the scaling dimension
$d_u$ of the vector unparticle operator and the new unparticle-SM
top quark couplings given in Eq. (1).

The scalar unparticle propagator is obtained with the help of scale
invariance and using the two point function of the unparticle
\cite{trece, dieciseis},

\begin{eqnarray}
\int\,d^4x\,
e^{ipx}\,<0|T\Big(O_U(x)\,O_U(0)\Big)0>&=&i\frac{A_{d_u}}{2\,\pi}\,
\int_0^{\infty}\,ds\,\frac{s^{d_u-2}}{p^2-s+i\epsilon} \nonumber
\\
&=&i\,\frac{A_{d_u}} {2\,sin\,(d_u\pi)}\,(-p^2-i\epsilon)^{d_u-2}
\, , \label{propagator}
\end{eqnarray}
with
\begin{eqnarray}
A_{d_u}=\frac{16\,\pi^{5/2}}{(2\,\pi)^{2\,d_u}}\,\frac{\Gamma(d_u+\frac{1}{2})}
{\Gamma(d_u-1)\,\Gamma(2\,d_u)} \, . \label{Adu}
\end{eqnarray}
The scale dimension $d_u$ is restricted in the range $1< d_u <2$.
Here,  $d_u>1$ is due to the non-integrable singularities in the
decay rate \cite{trece} and $d_u<2$ is due to the convergence of the
integrals \cite{dieciseis}.

The contribution induced by the scalar unparticle operator to the
CMDM (Fig. 1) is given by
\begin{eqnarray}
\Delta\kappa^{{\cal O}_U} &=&
\left(\frac{1}{\Lambda_u^{d_u-1}}\right)^2
\frac{A_{d_u}}{\sin(d_u\pi)}\frac{(2-d_u)}{4-d_u}
\frac{m_t^2}{8\pi^2} \nonumber \\
&\times & \int_0^1 dx \int_0^{1-x} dy \frac{(1-x-y)^{1-d_u}}{(m_t^2(x+y)^2)^{2-d_u}}\nonumber \\
&&(x+y)\left((-2+x+y)\zeta_S^2-(x+y)\zeta_A^2\right)
\end{eqnarray}
while the vector unparticle operator contribution to the CMDM is
\begin{eqnarray}
\Delta\kappa^{{\cal O}^\alpha_U} &=&
\left(\frac{1}{\Lambda_u^{d_u-1}}\right)^2
\frac{A_{d_u}}{\sin(d_u\pi)} \frac{m_t^2}{8\pi^2}
\int_0^1 dx \int_0^{1-x} dy
\frac{(1-x-y)^{1-d_u}}{(m_t^2(x+y)^2)^{2-d_u}}\nonumber \\
&\times&\left\{\left(\left[-2x(1-2y-x)-2y(1-y)\right]C_V^2\right. \right.\nonumber\\
&+&\left.\left[8+2x(-5+2y+x)-2y(5-y)\right]C_A^2 \right)\nonumber\\
&+&\frac{(1-x-y)}{(x+y)^2}\left(\left[x^2(x+1+y)+y^2(y+1+x)+xy\right]C_V^2\right.\nonumber\\
&+&\left.\left[x^2(x-1-5y)+y^2(1-y)+xy\right]C_A^2 \right)\nonumber\\
&+&\frac{4(1-x-y)}{2-d_u}\left(\left[3y^2+3x^2+2y+2x+4xy-2\right]C_V^2\right.\nonumber\\
&+&\left.\left.\left[3y^2+3x^2-4x+3xy+2\right]C_A^2\right)\right\}
\end{eqnarray}

The SM predicts, at the one loop level \cite{nueve}, $\Delta k^{SM}
= -0.056$. In order to get the respective predictions of the scalar
and vector unparticles to the CMDM, we have to add to the SM CMDM
the values obtained with the expressions given in Eqs. (5) and (6).
In Fig. 2 we depict the SM plus the respective scalar unparticle
contribution to the CMDM for different choices of the coefficients
$\zeta_{S,A}$. We have found that all three possible combinations
for the unparticle coefficients $C_{S,A}$ induce negative values for
the CMDM that are consistent with both bounds on $\Delta\kappa$
coming from the Tevatron and the $b\to sg$ decay measurements
\cite{nueve2, once}. On the other hand, we can appreciate in Fig. 3
that the latter bound already induces significant constraints on the
purely axial-vector ($C_V=0,C_A=1$) and quiral ($C_V=1,C_A=1$)
operator contributiuons to the top-quark CMDM, while the vector
contribution ($C_V=1,C_A=0$) satisfies both bounds except for low
values of the vector scale dimension $d_u$ of order $1.0-1.2$. A
similar sensitivity for small values of the scalar-unparticle
dimension $d_u$ has been also observed for the lepton-flavor
conserving Z boson decays \cite{quince}. In Fig.4 we have depicted
also the vector-unaprticle plus SM contributions to $\Delta\kappa$
for a higher value of the scale energy $\Lambda$. We can appreciate
that the constraints on the vector-unparticle contributions are not
very senstitive to this energy scale. However, for this energy scale
all three combinations of the vector-unparticle operators satisfy
both limits on $\Delta\kappa$, except for low values of the
respective scale dimension $d_u$.

In conclusion, we have studied the effect of scalar and vector
unparticle operators to the quark top CMDM at the one-loop level.
We have found that this dipole moment is sensitive to the values
of the respective unparticle dimensions $d_u$ and the low energy
scale $\Lambda$. We have used the bounds imposed on the top-quark
CMDM by the low-energy precision constraints \cite{nueve2} and
the Tevatron measurements \cite{once} in order to get limits on
the unparticle parameter $d_u$ and $\Lambda$. We would like to
mention that our limits on the vector unparticle dimension $d_u$
are similar to those obtained from FCNC processes involving a
quark-unparticle interaction in the context of CP-violation
effects in B decays \cite{catorce}, lepton-flavor conserving
decays of the Z boson \cite {dieciseis} and the current Tevatron
measurements for $t\bar t$ production\cite{quince}.\\

{\bf Acknowledgments}\\

This work is supported by Fundaci\'on del banco de la Republica
(Colombia), CONACYT     (M\'exico), CLAF and CONICET (Argentina).

\newpage

\begin{figure}
\includegraphics[height=10cm]{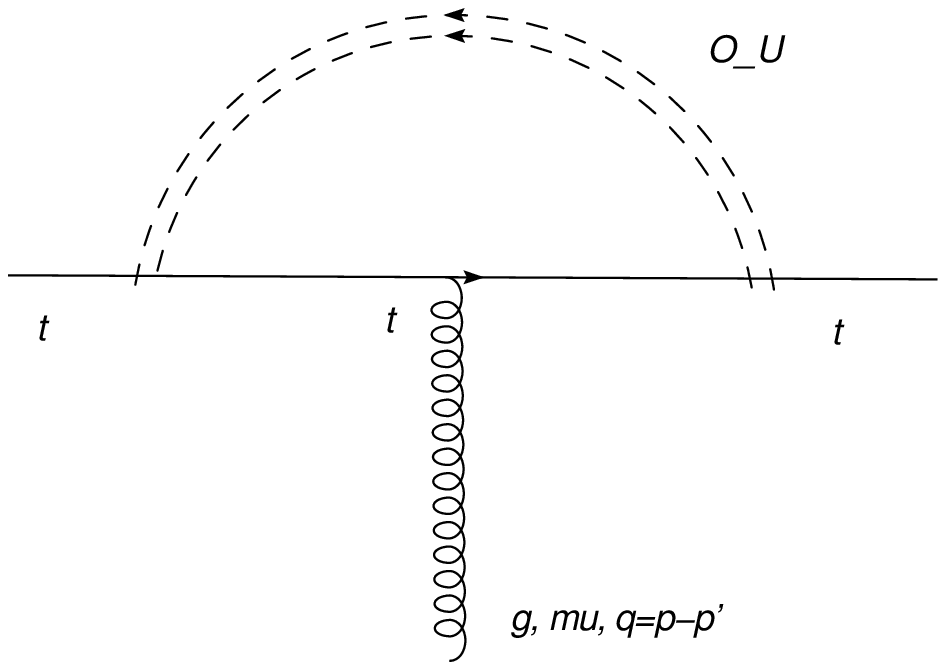}
\caption{Feynman diagrams for the contribution of the scalar and
vector unparticles to the CMDM of the top quark.}
\end{figure}

\newpage

\begin{figure}
\includegraphics[height=10cm,angle=-90]{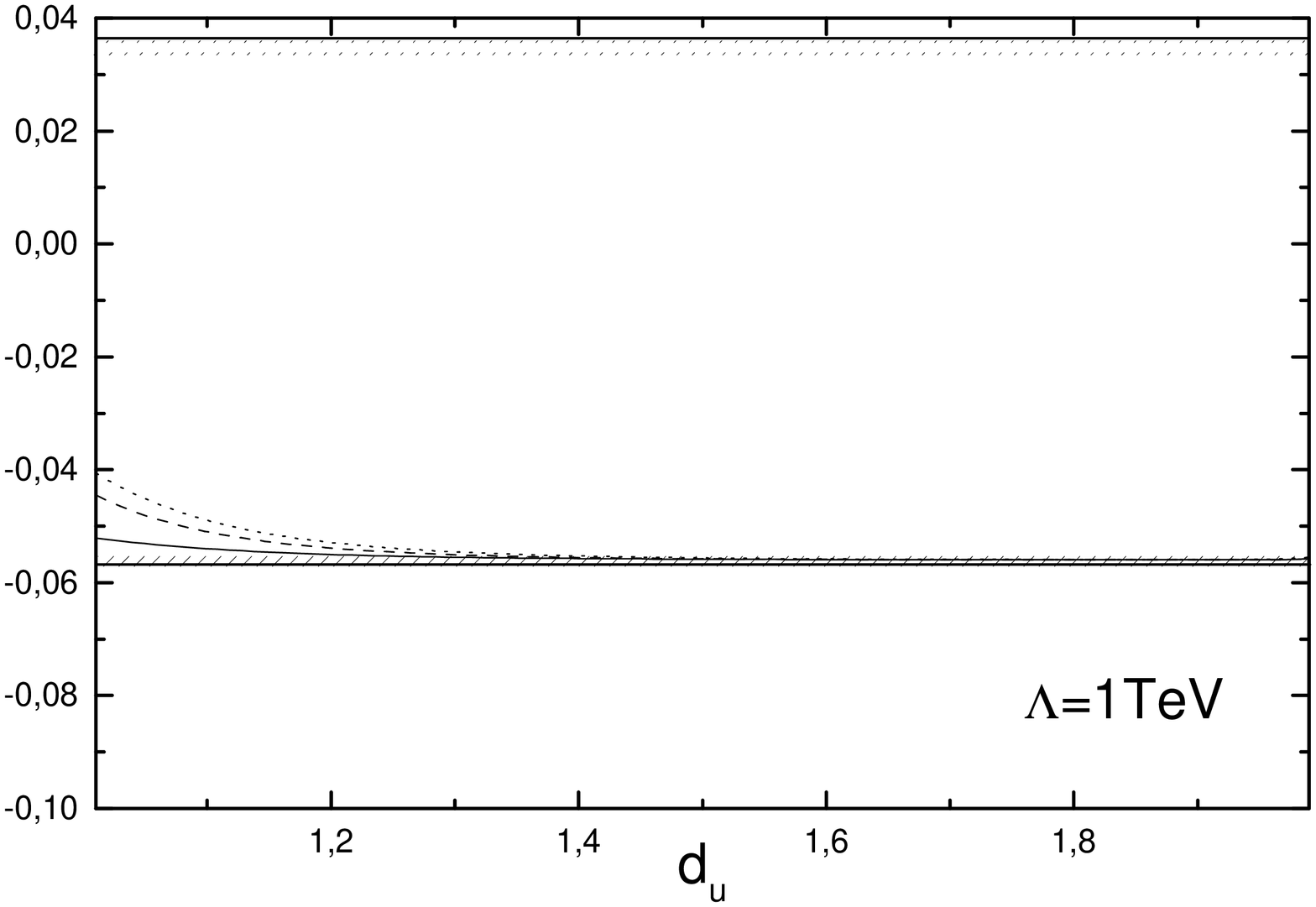}
\caption{The top-quark CMDM $\Delta\kappa$ as a function of the
scalar unparticle dimension $d_u$, with $\Lambda = 1 TeV$, for
purely scalar (solid line,  $\zeta_S = 1, \zeta_A = 0)$, purely
pseudoscalar (dashed, $\zeta_S = 0, \zeta_A = 1)$ and the quiral
combination (dotted, $\zeta_S = \zeta_A = 1)$ of the particle
operator. The solid stright lines give the limit induced on
$\Delta\kappa$ from the $b \to s\gamma$ decay at $95\%$ C.L..}
\end{figure}

\newpage

\begin{figure}
\includegraphics[height=10cm,angle=-90]{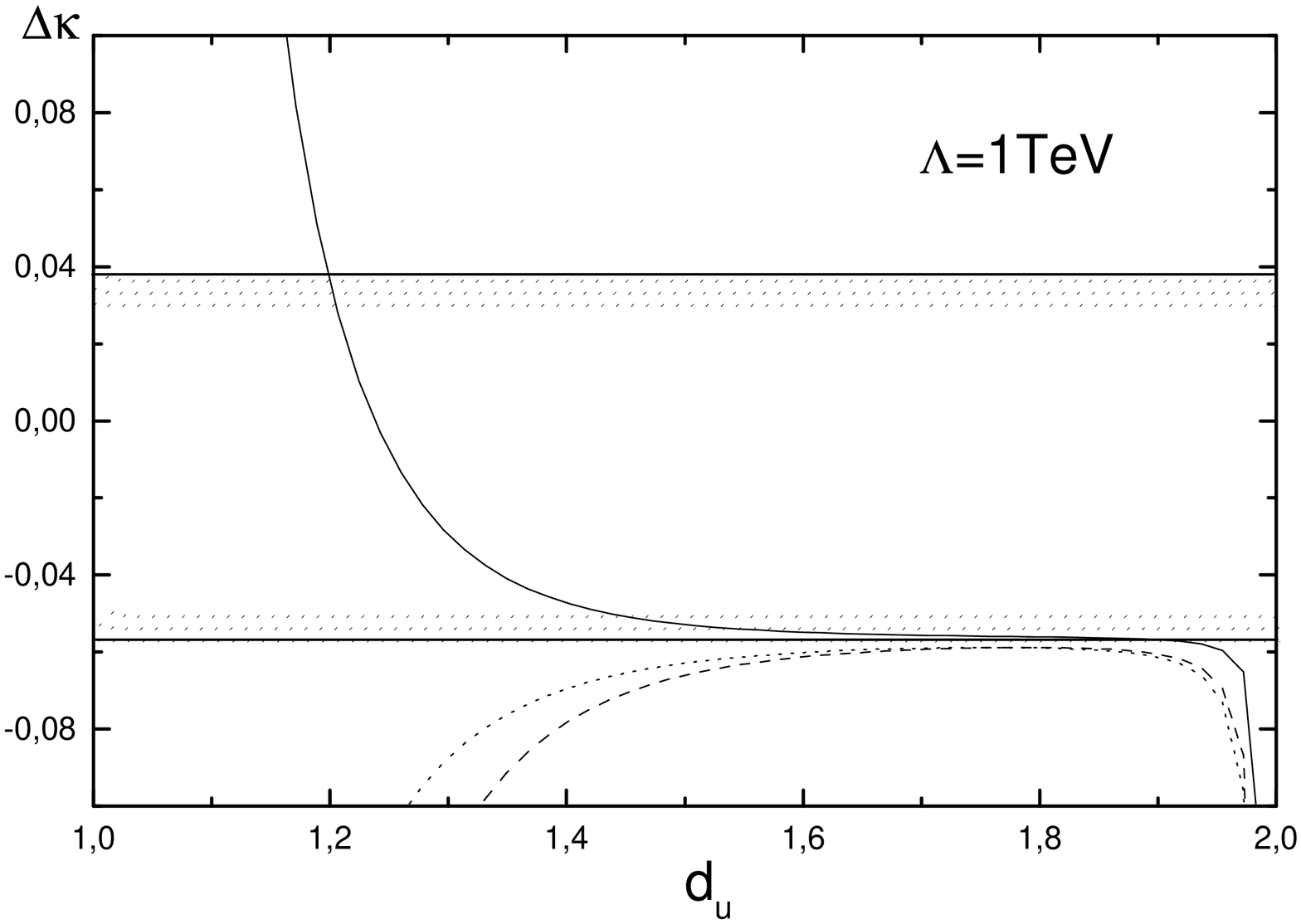}
\caption{ The CMDM of the top quark as a function of $d_u$ with
$\Lambda = 1$ for purely vector (solid line, $C_V = 1, C_A = 0)$,
axial vector (dashed $C_V =0, C_A = 1)$ and the quiral combination
(dotted, $C_V = C_A = 1)$ of the vector-unparticle operator. The
horizontal lines denote the limit obtained for $\Delta\kappa$ from
the $b \to s\gamma$ decay at $95\%$ C.L..}
\end{figure}

\newpage

\begin{figure}
\includegraphics[height=10cm,angle=-90]{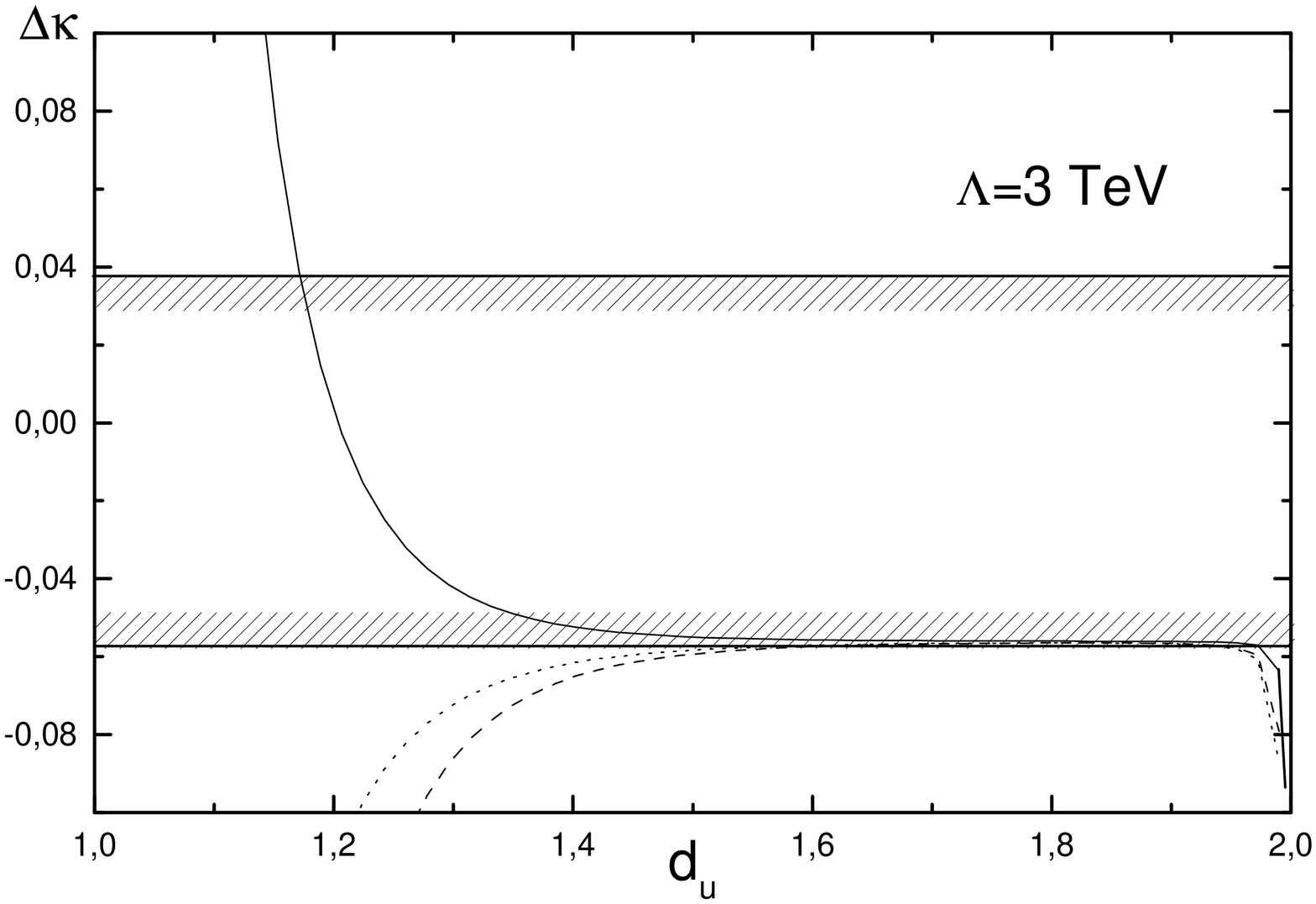}
\caption{Dependence of the vector-unparticle operator contributions
to the CMDM of the top quark for the scale energy $\Lambda=3$ TeV.
We use the same conventions as in Fig.3.}
\end{figure}

\end{document}